\documentclass[twocolumn,showkeys,preprintnumbers,amsmath,amssymb]{revtex4}
\usepackage{graphicx}
\usepackage{dcolumn}
\usepackage{bm}
\usepackage{epstopdf}
\usepackage{bm,hyperref,color,breakurl}

	\usepackage{fancyheadings}
\renewcommand{\thispagestyle}[1]{} 

\begin{document}


\vspace*{0.5cm}
\title{Nearest-neighbour antiferromagnetic interaction as a limiting factor\\ for critical temperature in model DMS system}


\author{K. Sza\l{}owski$\,$}
\email[]{kszalowski@uni.lodz.pl}
\affiliation{Department of Solid State Physics, Faculty of Physics and Applied Informatics,\\
University of \L\'{o}d\'{z}, ulica Pomorska 149/153, 90-236 \L\'{o}d\'{z}, Poland}



\begin{abstract}
In numerous diluted magnetic semiconductor (DMS) systems, the competition takes place between the short-range antiferromagnetic (AF) superexchange interactions and the long-range Ruderman-Kittel-Kasuya-Yosida (RKKY) coupling mediated by the charge carriers. Such a situation strongly influences the critical temperature, the maximization of which constitutes a challenging task in DMS physics and technology. The aim of the paper is to discuss theoretically the limiting effect of AF interactions between nearest-neighbour magnetic ions on the stability of inhomogeneous ferromagnetic state in a model diluted magnetic system reflecting some crucial features of DMS.

The modified molecular field-based model is constructed to account for the magnetic inhomogeneity. The behavior of the system is studied as a function of the ratio of superexchange integral to effective ferromagnetic coupling integral, including the possibility of clustering/anticlustering tendency for the magnetic ions. The ground state of the system is analysed. The critical temperature is found to change non-monotonically with the concentration of magnetic ions and decrease severely for larger concentrations. The behavior of the system significantly differs from the predictions of the usual homogeneous mean-field model. Brief comparison with selected experimental results for (Zn,Mn)Te is provided.
\end{abstract}

\keywords{diluted magnetic semiconductors, critical temperature, antiferromagnetic interactions, phase diagrams, ZnMnTe}

\maketitle

\section{Introduction}

Maximization of the critical temperature of DMS is of paramount importance from applicational point of view. One of the known obstacles is the presence of short-range superexchange interactions competing with the charge carrier-mediated RKKY interaction supporting the ferromagnetic state in those systems \cite{Dietl}. This results in significant reduction of effective magnetic moment and concentration of magnetically active impurity ions, which is observed experimentally. From the theoretical point of view, such systems are challenging due to the presence of magnetic frustration, positional disorder, and thus appearance of inhomogeneous magnetic orderings. Our aim is to construct a simple model of a disordered system with competing interactions (capturing some essential features met in DMS systems) and study its mean-field solution, focusing on the critical temperature.

\section{Model DMS system and its thermodynamics}

We consider a model site-diluted magnetic system consisting of spin $S$ impurity ions, distributed over the $N$ fcc lattice sites, with atomic concentration equal to $x$. This reflects the situation in typical DMS systems, where the substitutional ions of magnetic impurities occupy the sites of fcc (or slightly distorted fcc) lattice. The Hamiltonian of the model is the following: 
\begin{equation}\label{EqHamiltonian}
\mathcal{H}=-\frac{1}{2}\sum_{i,j}^{}{J\left(R_{ij}\right)\xi_i\,\mathbf{S}_i\,\xi_j\,\mathbf{S}_j}.
\end{equation}
The site dilution is introduced by the occupation number operators $\xi_{i}$, which take the value of 1 for a lattice site occupied by a magnetic impurity ion and 0 otherwise. The interaction between the spins is characterized by the exchange integral $J\left(R_{ij}\right)$. This interaction includes a long-range carrier-mediated RKKY coupling (driving the possible ferromagnetic order in DMS) and as well the antiferromagnetic (AF) interaction originating from superexchange mechanism, limited to nearest-neighbour (NN) magnetic ions \cite{Dietl}.

In a strongly diluted magnetic system, the dominant (or at least significant) number of impurity ions either lacks NN (single magnetic ions) or is involved in pairs without NN (isolated pairs) \cite{Behringer}. Therefore, the single magnetic ions experience only the long-range interaction, while the ions in pairs are additionally coupled antiferromagnetically. As a result, conditions for inhomogeneity in the magnetic ordering arise in the system. 

In order to take this feature into account in the thermodynamic description, we develop a modified molecular-field approximation (MFA) approach, proceeding along the lines sketched in our work \cite{Kilanski1}. The population of all magnetic impurities is divided into single ions and NN isolated pairs. The probability that a given impurity is an isolated ion equals to $p$. Therefore, the probability that it belongs to a NN pair reads $1-p$. All the clusters of larger size are neglected in the present approach.

Let us note that the lattice site occupations by magnetic impurities may not be purely random. The presence of an impurity in a given lattice site modifies the probability of occupying the NN sites, which then equals to $x\left(1+\alpha\right)$, where $\alpha$ is a Warren-Cowley parameter for NN, fulfilling some necessary inequalities \cite{Szalowski1}. Therefore, the probability $p$ reads in general $p=\left[1-x\left(1+\alpha\right)\right]^{12}$ 
for fcc lattice with 12 NN. 

MFA is known to work well in presence of the long-range interactions. Therefore, the density matrices describing the quantum states of single spins are assumed to take the MFA form (see e.g. \cite{Szalowski1}). 

For the spin of a single ion we use the density matrix 
\begin{equation}
\pmb{\rho}^{(1)}=\frac{1}{\mathcal{Z}^{(1)}} \,\exp\left(\frac{\lambda^{(1)}}{k_{\rm B}T}    \,\mathbf{S}^z\right),\,\,\mathcal{Z}^{(1)}=\mathrm{Tr}  \exp\left(\frac{\lambda^{(1)}}{k_{\rm B}T}\right),
\end{equation}
depending on the molecular field parameter $\lambda^{(1)}$ and yielding the magnetization $m^{(1)}=S\mathcal{B}_{S}\left(S\frac{\lambda^{(1)}}{k_{\rm B}T}\right)$. Here, $T$ is the absolute temperature, $k_{\rm B}$ is the Boltzmann constant and $S\mathcal{B}_{S}$ is a Brillouin function for spin $S$. For the ions belonging to NN pairs we accept the density matrices
\begin{equation}
\pmb{\rho}^{(2)}_{\pm}=\frac{1}{\mathcal{Z}^{(2)}_{\pm}} \,\exp\left(\frac{\lambda^{(2)}_{\pm}}{k_{\rm B}T}    \,\mathbf{S}^z\right),\,\,\mathcal{Z}^{(2)}_{\pm}=\mathrm{Tr}  \exp\left(\frac{\lambda^{(2)}_{\pm}}{k_{\rm B}T}\right)
\end{equation}
with the parameters $\lambda^{(2)}_{\pm}$ and $m^{(2)}_{\pm}=S\mathcal{B}_{S}\left(S\frac{\lambda^{(2)}_{\pm}}{k_{\rm B}T}\right)$, respectively. The signs $+$ and $-$ allow to distinguish between the non-equivalent ions in a pair. The total magnetization of a pair reads $m^{(2)}=m^{(2)}_{+}+m^{(2)}_{-}$. The total magnetization of the system per lattice site is a sum of contributions from the single ions and the pairs, namely $\overline{m}=\left\langle\left\langle \mathbf{S}^z\right\rangle\right\rangle_{r}=p\,m^{(1)}+\left(1-p\right)m^{(2)}/2$. 

Let us denote by $\left\langle \cdots \right\rangle$ the thermodynamic average over a canonical ensemble, while the configurational average over atomic disorder is $\left\langle \cdots \right\rangle_{r}$. Calculation of the thermodynamic and configurational average of the Hamiltonian (\ref{EqHamiltonian}) is performed by using the postulated density matrices. The way of calculating the two-site configurational averages over the disorder reflects the inhomogeneity in the system. For the sites $i$ and $j$ being NN, we assume $\left\langle\left\langle \xi_{i}\,\mathbf{S}_{i}\,\xi_{j}\,\mathbf{S}_{j}\right\rangle\right\rangle_{r}=x\left(1-p\right)m^{(2)}_{+}m^{(2)}_{-}/2$. For a pair of non-NN sites $\left\langle\left\langle \xi_{i}\,\mathbf{S}_{i}\,\xi_{j}\,\mathbf{S}_{j}\right\rangle\right\rangle_{r} = x^2\left[p^2 \left(m^{(1)}\right)^2\right.\newline  +\left.p\left(1-p\right)\,m^{(1)}m^{(2)} /2 +\left(1-p\right)^2\,\left(m^{(2)}\right)^2/4\right]$.

The internal energy of the system $U=\left\langle\left\langle \mathcal{H}\right\rangle\right\rangle_{r}$ yields: 
\begin{align}\label{EqEnthalpy}
&\frac{U}{N} = -x^2J_{F}\Biggl[\frac{p^2}{2} \left(m^{(1)}\right)^2 +\frac{p\left(1-p\right)}{2}\,m^{(1)}m^{(2)}\Biggr.\nonumber\\
&\Biggl.+\,\frac{\left(1-p\right)^2}{8}\left(m^{(2)}\right)^2\Biggr]-\frac{1-p}{2}\,xJ_{AF}\,m^{(2)}_{+}m^{(2)}_{-}.
\end{align}

The parameter $J_{F}=\sum_{k=2}^{\infty}{z_{k}\,J^{\rm RKKY}\left(r_{k}\right)}$ characterizes in a convenient way the long-range RKKY interaction, which is assumed to yield the net ferromagnetic coupling $J_{F}>0$. Here, $z_{k}$ denotes the number of sites for $k$-th coordination zone in the fcc lattice, having the radius of $r_{k}$. 

The entropy can be expressed as:
\begin{align}\label{EqEntropy}
&\frac{\mathcal{S}}{N}=x\Biggl\{ k_{\rm B}\Biggl[p\,\ln \mathcal{Z}^{(1)}+\frac{1-p}{2}\left(\ln \mathcal{Z}^{(2)}_{+}+\ln \mathcal{Z}^{(2)}_{-}\right)\Biggr]\Biggr.\nonumber\\
&-\Biggl.\frac{1}{T}\left[p\lambda^{(1)}m^{(1)}+\frac{1-p}{2}\left(\lambda^{(2)}_{+}m^{(2)}_{+}+\lambda^{(2)}_{-}m^{(2)}_{-}\right)\right]\Biggr\}.
\end{align}
Knowing the above, the Helmholtz free energy can be calculated as $F=U - \mathcal{S}T$. From the variational minimization of the Helmholtz energy with respect to the three molecular field parameters, the following set of equations is obtained:
\begin{align}\label{EqEquationsforLambda}
\lambda^{(1)}&=\left(pm^{(1)}+\frac{1-p}{2}\,m^{(2)}\right)xJ_{F}\nonumber\\
\lambda^{(2)}_{+}&=J_{AF}m^{(2)}_{-} +\left(pm^{(1)}+\frac{1-p}{2}\,m^{(2)}\right) xJ_{F}\nonumber\\
\lambda^{(2)}_{-}&=J_{AF}m^{(2)}_{+} +\left(pm^{(1)}+\frac{1-p}{2}\,m^{(2)}\right) xJ_{F}.
\end{align}
Solving the equations self-consistently allows to find the magnetizations of all the subsystems.

\subsection{Ground state}
\begin{figure}[h]%
\includegraphics*[width=\linewidth]{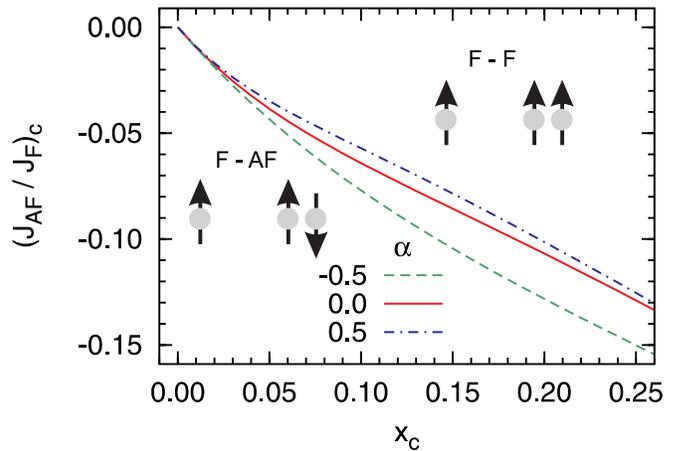}
\caption{%
  The ground-state phase diagram of the model as dependent on the concentration of magnetic impurities $x$ and the relative strength of AF coupling $J_{AF}/J_{F}$ for various values of a Warren-Cowley parameter for NN, describing the non-randomness in the distribution of impurities.}
\label{FigGroundState}
\end{figure}

\begin{figure*}[t]%
  \includegraphics*[width=.80\textwidth]{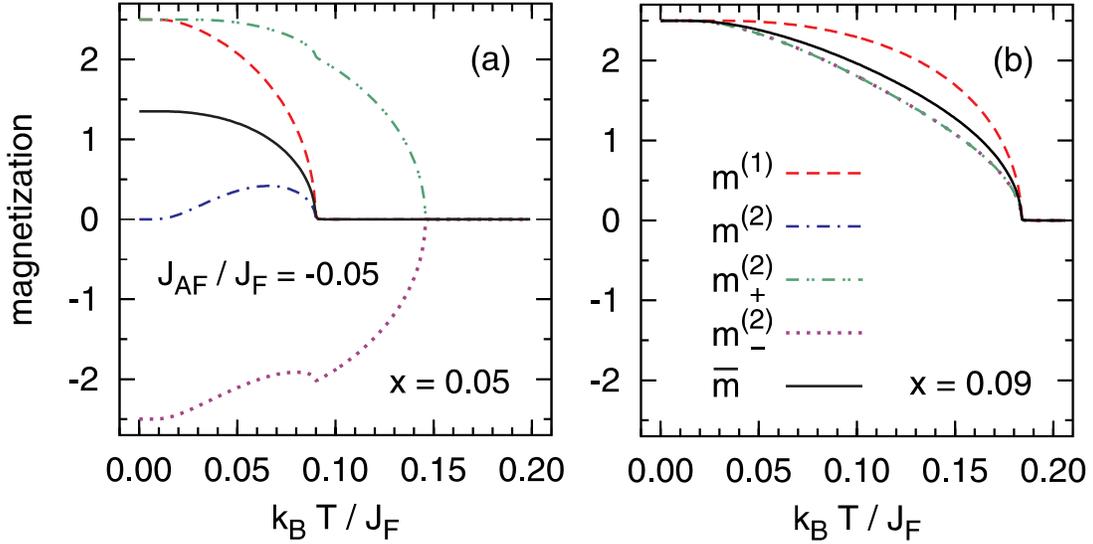}%
  \caption[]{%
    The temperature dependencies of the magnetization components for $J_{AF}/J_{F}=-0.05$, for the magnetic impurity concentration $x=0.05$ corresponding to the phase F-AF (a) and $x=0.09$ corresponding to the phase F-F (b).}
    \label{FigTemperatureMagnetization}
\end{figure*}

The ground-state of the system (at the temperature $T=0$) can be determined basing on the minimization of the internal energy given by Eq. (\ref{EqEnthalpy}). Such a procedure shows the existence of two possible magnetic orderings: a phase with the ferromagnetic order of the single ions and the antiferromagnetic order within the pairs of ions (F-AF) and a frustrated phase with all the spins ordered ferromagnetically (F-F), depending on the relative strength of antiferromagnetic and ferromagnetic couplings $J_{AF}/J_{F}$ and the concentration of magnetic component $x$. The phase boundary follows from the comparison of the internal energies of both phases and is described by the expression
\begin{equation}
\left(J_{AF}/J_{F}\right)_{c}=-\frac{1}{2}\,x_{c}\left\{1+\left[1-x_{c}\left(1+\alpha\right)\right]^{12}\right\}.
\end{equation}
In the Fig.~\ref{FigGroundState} the stability ranges of both phases are illustrated for purely random distribution of the magnetic ions as well as for the clustering ($\alpha>0$) and anticlustering ($\alpha<0$) tendency. Let us note that the stability range of F-F phase may be expected to shift toward larger concentrations $x$ when the larger clusters of AF-coupled ions are included in the model. 

\subsection{Temperature magnetization curves}

By solving numerically the self-consistent set of equations (\ref{EqEquationsforLambda}) together with the relations between magnetizations and molecular field parameters following from the choice of the density matrices, the temperature dependence of the magnetization can be found. Since the system of equations has in general more than one solution, it is necessary to select the one consistent with the ground state (discussed above). All the further calculations assume $S=5/2$.

Fig.~2 presents the example calculations performed for a rather weak AF interactions, $J_{AF}/J_{F}=-0.05$. For the lower concentration (corresponding to F-AF state), an interesting feature is that the increase of the temperature causes the appearance of net non-zero magnetization $m^{(2)}$ of the ion pairs. The magnetization of the single ion system $m^{(1)}$ vanishes at certain temperature $T_{c}$, together with the net magnetization of the pairs $m^{(2)}$ and hence the total magnetization of the system $\overline{m}$. However, $m^{(2)}_{\pm}$ do not vanish itself and the spins of ions in pairs remain antiferromagnetically polarized up to the N\'{e}el temperature $T_{N}>T_{c}$. The situation is quantitatively different for the larger concentration $x$, corresponding to F-F ground state. As visible in Fig.~2, both the single ions and the pairs are ferromagnetically polarized up to the critical temperature $T_{c}$ and the magnetizations of all the subsystems vanish simultaneously.  

\subsection{Critical temperature}
For the homogeneous ferromagnetic state (i.e. for $J_{AF}=0$) we get the reference critical temperature $T_{c}=S\left(S+1\right)xJ_{F}/3k_{\rm B}$. On the other hand, in the limit of a rather strong superexchange we deal with two completely decoupled systems and then only the single magnetic ions contribute to the ferromagnetic state, leading to the critical temperature of $T_{c}=S\left(S+1\right)x\,pJ_{F}/3k_{\rm B}$.

\begin{figure}[t]%
\includegraphics*[width=\linewidth]{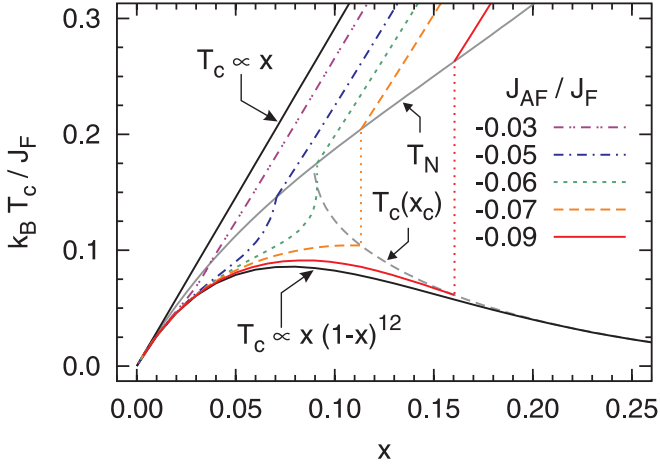}
\caption{%
  The normalized critical temperature (at which the total magnetization of the system vanishes) as a function of the magnetic impurities concentration $x$, for various relative magnitudes of AF coupling $J_{AF}/J_{F}$ and for randomly distributed impurities.}
\label{FigCriticalTemperature}
\end{figure}

The usual procedure of finding the critical temperature within MFA approach relies on the linearization of the relations between magnetization and molecular field parameters resulting from the choice of the density matrices \cite{Szalowski1}. It is tempting to apply this procedure here, however it should be preceded by the careful analysis of the temperature magnetization curves, since the magnetizations of the subsystems in the F-AF phase do not vanish simultaneously. As a consequence, the critical temperature $T_{c}$ should be determined numerically on the basis of the temperature magnetization curves. As for $T>T_{c}$ the magnetization $m^{(1)}$ is zero, then for finding the N\'{e}el temperature the usual linearization procedure can be utilized, leading to the result $T_{N}=S(S+1)\left|J_{AF}\right|/3k_{\rm B}$, being independent of the existence of the ferromagnetic interactions in the system. On the other hand, for F-F phase, all the components of magnetization vanish together and the linearization procedure gives 
\begin{align}\label{EqTc}
T_{c}=&\frac{S\left(S+1\right)}{6 k_{\rm B}} \bigg[ \sqrt{x^2J_{F}^2+2xJ_{F}J_{AF}\left(1-2p\right)+J_{AF}^2}\bigg.\nonumber\\
&+\bigg. xJ_{F}+J_{AF}\bigg].
\end{align}

Fig.~(\ref{FigCriticalTemperature}) presents the dependence of the normalized critical temperature $k_BT_c/J_F$ (at which the total magnetizaton of the system vanishes) on the atomic concentration of the magnetic component $x$. A purely random, uncorrelated distribution of the magnetic impurity ions over the lattice is assumed ($\alpha=0$). The plot is prepared for varying relative strength of AF interaction, the figure of merit being the ratio of NN antiferromagnetic exchange constant $J_{AF}$ to the ferromagnetic long-range interaction parameter $J_F$. Both the asymptotes - the homogeneous MFA result neglecting the existence of AF interaction and the limiting behavior of $T_{c}$ for strong $J_{AF}$ - are shown with the bold lines. It is visible that the presence of $J_{AF}$ always lowers the critical temperature below its homogeneous MFA value. Within the range of F-AF phase, for weak $J_{AF}$ the critical temperature increases monotonically with $x$. Increasing the strength of AF interaction causes the critical temperature to exhibit a local maximum for certain value of $x$ and then significantly drop. On the other hand, for F-F phase, $T_{c}$ changes almost perfectly linearly with $x$, however its value is reduced over a constant depending on $J_{AF}$. It is noticeable that for $J_{AF}/J_{F}<-0.059$ the critical temperature exhibits a jump with increasing $x$, between F-AF and F-F phase (as shown by the dotted vertical lines). The envelope of the maximum critical temperatures $T_{c}\left(x_{c}\right)$ achieved in the F-AF phase within this range is also shown. 

\begin{figure}[t]%
\includegraphics*[width=\linewidth]{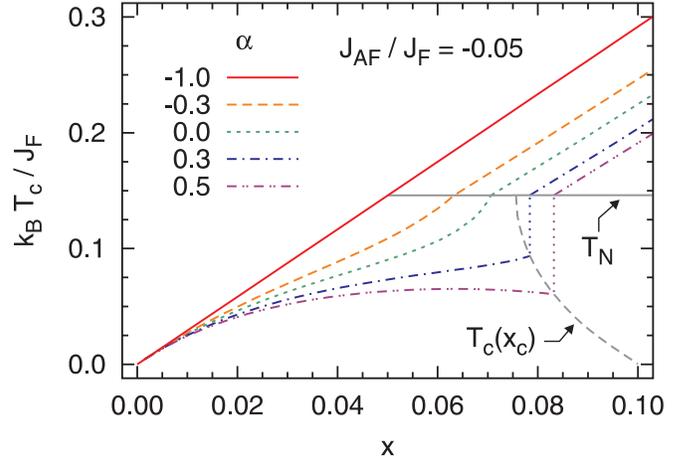}
\caption{%
  The normalized critical temperature as a function of the magnetic impurities concentration $x$, for $J_{AF}/J_{F}=-0.05$ and for various values of the Warren-Cowley parameter for NN.}
\label{FigWarrenCowley}
\end{figure}

The influence of the clustering or anticlustering tendency of magnetic ions is analysed for the weak $J_{AF}$ in the Fig.~\ref{FigWarrenCowley}. It can be observed that the effect of clustering ($\alpha>0$) limits the $T_{c}$ value and is similar to the result of increasing the strength of AF coupling. On the contrary, the tendency of anticlustering ($\alpha<0$) reduces the influence of AF interaction, and in the limiting case of $\alpha=-1$ (without NN) restores the ordinary homogeneous MFA result for $T_{c}$. This indicates the significant importance of the possible non-randomness of the magnetic ions distribution in DMS in presence of the superexchange.

\begin{figure*}[t]%
  \includegraphics*[width=.80\textwidth]{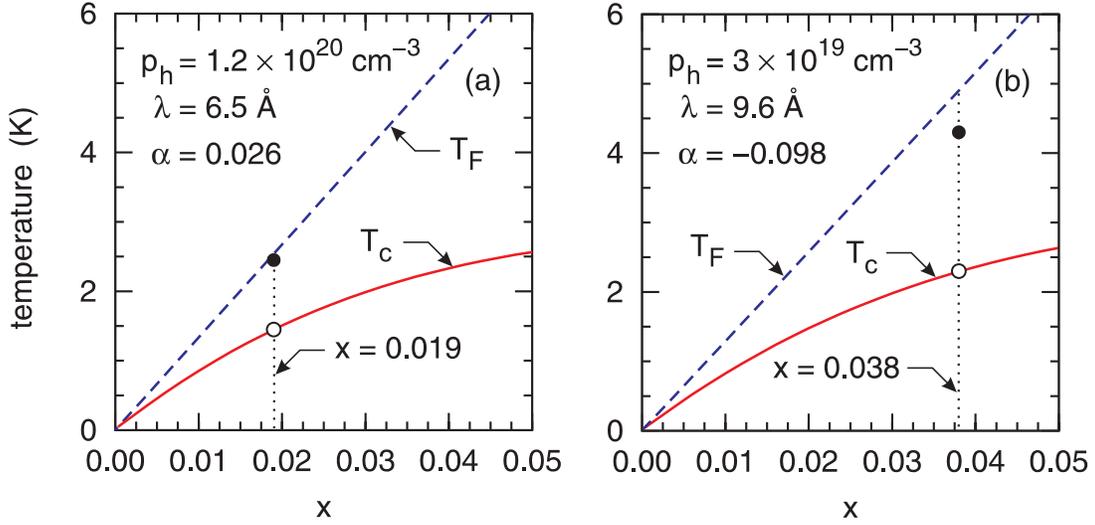}%
  \caption[]{%
    The critical temperature of p-doped (Zn,Mn)Te, for $p_{h}=1.2\times 10^{20}$ cm$^{-3}$ (a) and $p_{h}=3\times 10^{19}$ cm$^{-3}$ (b). The predictions of our model are shown with solid lines (critical temperature) and dashed lines (characteristic ferromagnetic temperature). The corresponding experimental values (after \cite{Ferrand1}) are marked with empty and filled circles, respectively.}
    \label{FigExp}
\end{figure*}

\section{Comparison with experimental data for (Z\lowercase{n},M\lowercase{n})T\lowercase{e}}

It might be instructive to present briefly some results of application of our model to a representative DMS system. For this purpose we selected the p-doped Zn$_{1-x}$Mn$_{x}$Te, a II-VI semiconductor exhibiting charge carriers-driven ferromagnetism, which has been subject to experimental and theoretical studies of Ferrand \emph{et al.} \cite{Ferrand1}. 

In order to calculate the parameter $J_{F}$, we make use of the RKKY exchange integral in the following form:
\begin{align}
J^{RKKY}\left(r\right)=&\frac{m_{h}a^6 \left(N_0\beta\right)^2}{8\pi h}k_{\rm F}^4\,e^{-r/\lambda}\nonumber\\
&\times \frac{\sin \left(2 k_{\rm F}r\right)-\left(2 k_{\rm F}r\right)\cos\left(2 k_{\rm F}r\right)}{\left(2 k_{\rm F}r\right)^4}.
\end{align}
After the work \cite{Ferrand1}, we assume the interaction to be mediated dominantly by the heavy holes of effective mass $m_{h}=0.60\,m_{e}$, therefore $k_{\rm F}=\left(3\pi^2 p_{h}\right)^{1/3}$ is a wavevector for the heavy holes in the valence band. The spin-hole exchange integral $N_0\beta=-1.1$ eV and the lattice constant $a=6.1$ $\mathrm{\AA}$ are accepted. The limited mean free path of the holes in a disordered system is reflected by an exponential damping of RKKY coupling with a characteristic length $\lambda$. 

The dominant AF superexchange integral for the NN in (Zn,Mn)Te equals to $J_{AF}=-9$ K (after the Ref.~\cite{Bindilatti1}). However, the same experimental results indicate that the AF interaction between Mn impurities in this DMS is not limited to NN ions. Thus, the exchange integrals $J^{AF}\left(r_{k}\right)$, known up to the distance of the fifth neighbors, are included in the calculation of $J_{F}$, so that finally $J_{F}=\sum_{k=2}^{\infty}{z_{k}\,\left[J^{\rm RKKY}\left(r_{k}\right)+J^{AF}\left(r_{k}\right)\right]}$. 

In the Fig.~\ref{FigExp} we present the comparison of our model predictions with the experimental data for two representative samples studied in the Ref.~\cite{Ferrand1}. For both samples the Curie-Weiss temperature (close to the critical temperature $T_{c}$) was measured and the temperature $T_{F}$ characterizing only the ferromagnetic charge carriers-driven interactions was evaluated. In the model of Ferrand \emph{et al.} is was done using the empirical characteristic temperature $T_{AF}$, determined from the experiments on the samples with sufficiently low concentrations of charge carriers. In the Fig.~\ref{FigExp} the temperatures $T_{c}$ and $T_{F}$ taken from \cite{Ferrand1} are marked by empty and full circles, respectively. The experimentally determined concentrations of Mn impurities $x$ and holes $p_{h}$ are indicated in the plots. For both samples also the effective concentration of magnetic impurities was obtained from the low-temperature saturation magnetization, what allowed us to establish the value of $p$ and thus calculate the Warren-Cowley parameter $\alpha$ necessary to reproduce $p$ for the given concentration $x$. 

Using the above-mentioned parameters, we reproduce the experimentally observed critical temperatures using the characteristic damping lengths $\lambda$ slightly larger than the lattice constant, as given in the plots. Moreover, the same set of parameters allows us to calculate the characteristic ferromagnetic temperature $T_{F}=xS\left(S+1\right)\sum_{k=2}^{\infty}{z_{k}\,J^{\rm RKKY}\left(r_{k}\right)}/3$ (which does not include AF superexchange interactions, to conform with the meaning of experimentally determined $T_{F}$ in the work \cite{Ferrand1}). The values of $T_{c}$ and $T_{F}$ were calculated numerically as a function of magnetic impurity concentration (assuming constant $\alpha$). 

It can be observed that for the sample with lower Mn concentration $x=0.019$ (Fig.~\ref{FigExp}(a)), we reproduce well both the value of the critical temperature and the characteristic ferromagnetic temperature within our model. On the other hand, for the sample with higher Mn concentration $x=0.038$ (Fig.~\ref{FigExp}(b)), the value of $T_{F}$ is slightly overestimated. Such a difference might be attributed to the influence of larger clusters of the magnetic ions, which is more pronounced for larger $x$. Especially, the clusters containing odd number of the ions contribute to the ground-state magnetization (contrary to the AF-polarized pairs of ions) and thus have an impact on the value of $p$ extracted from the experimental data. 

Let us emphasize that in our model we do not make use of the empirically determined characteristic temperature $T_{AF}$, but we use only the values of superexchange and RKKY exchange integrals present in the Hamiltonian (\ref{EqHamiltonian}).

\section{Conclusions}

For our model DMS system, it is noticeable that the critical temperature is rather sensitive to the presence and the relative strength of the AF interactions between NN as well as to the clustering or anticlustering of the magnetic ions. Application of the model to (Zn,Mn)Te indicates that the experimental results can be satisfactorily reproduced, especially for low concentrations of magnetic impurities, by constructing the systematic thermodynamic treatment of the system Hamiltonian.

The future developments of the model may especially include taking into account the existence of the larger clusters of magnetic ions (e.g. triples), since the relative abundance of such clusters rises rapidly with the concentration of impurities. It may also be worthwhile to include the AF interactions between further neighbors in the magnetic sublattice. 

\section*{Acknowledgement}
Prof. T. Balcerzak is gratefully acknowledged for fruitful discussions and critical reading of the manuscript.

%

\begin{thebibliography}{[1]}

\bibitem{Dietl}%
 T.~Dietl,
 in: Handbook of Magnetism and Advanced Magnetic Materials vol. 5, edited by H.~Kronm\"{u}ller and S.~Parkin (J. Wiley \& Sons, 2007), pp. 2774-2789.
 
\bibitem{Behringer}%
 R.\,E.~Behringer, J. Chem. Phys. \textbf{29}, 537 (1958).

\bibitem{Kilanski1}%
L.~Kilanski, R.~Szymczak, W.~Dobrowolski, K.~Sza{\l}owski, V.\,E.~Slynko, and E.\,I.~Slynko, Phys. Rev. B \textbf{82}, 094427 (2010).
 
\bibitem{Szalowski1}%
 K.~Sza{\l}owski, and T.~Balcerzak, Phys. Rev. B \textbf{77}, 115204 (2008).
 
 \bibitem{Ferrand1}%
D.~Ferrand, J.~Cibert, A.~Wasiela, C.~Bourgognon, S.~Tatarenko, G.~Fishman, T.~Andrearczyk, J.~Jaroszy\'nski, S.~Kole\'snik, T.~Dietl, B.~Barbara, and D.~Dufeu, Phys. Rev. B \textbf{63}, 085201 (2001).

\bibitem{Bindilatti1} %
V.~Bindilatti, E.~ter~Haar, N.~F.~Oliveira,~Jr., Y.~Shapira, and M.~T.~Liu, Phys. Rev. Lett. \textbf{80}, 5425 (1998).

 \end{thebibliography}
%

\end{document}